\documentclass{appolb}
\usepackage{amsmath,amssymb,mathrsfs,graphicx,slashed,ragged2e}

\newcommand{\Sig}{\Sigma_{1}}
\newcommand{\Sigb}{\overbar{\Sigma}_{1}}

\newcommand{\mb}{\overbar{m}}
\renewcommand\thesection{\Roman{section}}
\renewcommand\thesubsection{\Alph{subsection}}

\newcommand{\overbar}[1]{\mkern 1.5mu\overline{\mkern-1.5mu#1\mkern-1.5mu}\mkern 1.5mu}

\newcommand\mtiny[1]{\mbox{\tiny\ensuremath{#1}}}

\newcommand{\Rh}{\overset{\mtiny{h}}R{\vphantom{R}}}
\newcommand{\Rg}{\overset{\mtiny{g}}R{\vphantom{R}}}

\newcommand{\Th}{\overset{\mtiny{h}}T{\vphantom{T}}}

\newcommand{\Gammah}{\overset{\mtiny{h}}\Gamma{\vphantom{\Gamma}}}

\newcommand{\psih}{\overset{\mtiny{h}}\psi{\vphantom{\psi}}}

\newcommand{\nablag}{\overset{\mtiny{g}}\nabla{\vphantom{\nabla}}}
\newcommand{\squareg}{\overset{\mtiny{h}}\square{\vphantom{\square}}}
\renewcommand{\squareg}{\overset{\mtiny{g}}\square{\vphantom{\square}}}

\begin{document}
\title{Signatures of Extended Theories of Gravity in Black Hole Oscillations%
\thanks{Presented at the 6th Conference of the Polish Society on Relativity.}%
}
\author{Arthur George Suvorov\footnote{arthur.suvorov@tat.uni-tuebingen.de}
\address{Theoretical Astrophysics, IAAT, University of T{\"u}bingen, Germany}
\\
}
\maketitle
\begin{abstract}
In general relativity, the Kerr metric uniquely represents the geometry surrounding an isolated, rotating black hole. An identification of significant non-Kerr features in some astrophysical source would then provide a `smoking-gun' for the break-down of general relativity in the strong-field regime. On the other hand, Kerr black holes are common to many other theories of gravity, and thus a validation of the Kerr metric does not necessarily favour general relativity amongst all possibilities. The nature of gravitational perturbations will however differ between different theories of gravity. Future precision tests involving gravitational waves from oscillating black holes, such as identifications of the quasi-normal mode spectrum from ring-down, will thus be able to probe the underlying theory, even if the object is Kerr. Here, we write down the equations governing metric perturbations of a Kerr black hole in $f(R)$ gravity in a form that is more conducive to numerical study.
\end{abstract}
\PACS{04.30.-w, 04.50.Kd, 04.70.Bw}
  
\section{Introduction}

Many higher-order curvature theories of gravity, such as the $f(R)$ and Gauss-Bonnet theories, are designed so that they reduce to general relativity (GR) in some appropriate limit. In order to be viable, a given theory must, in addition to respecting weak-field demands coming from solar system experiments, abide by various strong-field constraints set by compact objects. Observations in the electromagnetic spectrum which probe the nature of black holes (e.g. quasiperiodic oscillations of microquasars) suggest that they are, geometrically speaking, consistent with the (unique) Kerr solution of GR \cite{kerrmetric}, though there is room for modified gravity parameters to be non-zero \cite{kerrtest,kerrtest2}. In any event, since these higher-order theories contain GR as a limiting case, they necessarily contain the Kerr solution under some circumstances \cite{psaltis}. Therefore, a validation of the Kerr spacetime as a description for astrophysical black holes does not necessarily favour GR amongst all possibilities \cite{psaltis}.

Nevertheless, with the advent of gravitational wave astronomy, additional tests of black hole character are becoming available. Indeed, the behaviour of gravitational perturbations is inherently tied to the structure of the field equations \cite{bar08}. A study of the oscillation spectrum of black holes in extended theories of gravity, even if Kerr, can thus pave the way for future precision tests of GR \cite{perts}. Recently, we developed a technique to write down a decoupled set of (wave) equations governing metric perturbations of a Kerr black hole in $f(R)$ gravity \cite{suv19}. The equations presented in \cite{suv19} were however written down using the compact notation of the Newman-Penrose (NP) formalism \cite{np62}, and thus may not be readily usable. Here, we write down these equations in coordinate form, so that they are better suited to numerical analysis. The explicit appearance of $f(R)$ gravity terms becomes apparent in this way.

\section{Field equations}

In the $f(R)$ class of theories, the Ricci scalar, $R$, defining the Einstein-Hilbert Lagrangian, is replaced by an arbitrary function of this quantity, $f(R)$. The field equations read\footnote{Throughout this work, we use natural units with $G=c=1$ and adopt a time-like $(+,-,-,-)$ metric signature. Complex conjugation is indicated by an overhead bar.} 
\begin{equation} \label{eq:fofr1}
f'(R) R_{\mu \nu} -  \frac {f(R)} {2} g_{\mu \nu} + \left( g_{\mu \nu} \square - \nabla_{\mu} \nabla_{\nu} \right) f'(R) = 8 \pi T_{\mu \nu},
\end{equation}
where $R_{\mu \nu} = R^{\alpha}_{\mu \alpha \nu}$ is the Ricci tensor, $g_{\mu \nu}$ is the metric tensor, and $\square = \nabla_{\mu} \nabla^{\mu}$ denotes the Laplace-Beltrami operator. Setting $f(R) = R$ returns the Einstein equations, as expected.  Taking the trace of \eqref{eq:fofr1} yields a constraint between the Ricci scalar and the function $f$,
\begin{equation} \label{eq:treqn}
3 \square f'(R) + R f'(R) -2 f(R) = 8 \pi T^{\mu}_{\mu}.
\end{equation}

\subsection{Gravitational Perturbations}

Consider a background metric $\boldsymbol{g}$ which induces a vanishing Ricci tensor. As is easily verified, this metric solves the field equations \eqref{eq:fofr1} in vacuo provided that $f(0) = 0$. Through a slight abuse of notation, we introduce a perturbation $\boldsymbol{h}$ via
\begin{equation} \label{eq:mettpt}
g_{\mu \nu} \rightarrow g_{\mu \nu} + h_{\mu \nu}.
\end{equation}
Denoting perturbed quantities with an overhead $h$ and background quantities with an overhead $g$, the perturbation \eqref{eq:mettpt} induces a shift into the function $f$ through
\begin{equation} \label{eq:leadingf}
f(R) \rightarrow f(\Rg) + \Rh f'(\Rg),
\end{equation}
to leading order in $h$. As a result, if we assume that $f$ is an analytic function, then, because $f(\Rg) = f(0) = 0$, only quadratic corrections $a_{2}$ to the function $f$, viz. 
\begin{equation}
f(R) = R + \frac {a_{2}} {2} R^2 + \cdots,
\end{equation}
will appear within the perturbed field equations \eqref{eq:fofr1}. In general, a perturbation \eqref{eq:mettpt} in expression \eqref{eq:fofr1} leads to the system
\begin{equation} \label{eq:generalsystem}
\begin{aligned}
&f''(\Rg) \Rh \Rg_{\mu \nu} + f'(\Rg) \Rh_{\mu \nu} + g_{\mu \nu} \left\{ - g^{\beta \sigma} \Gammah^\alpha{}_{\beta \sigma} \nablag_{\alpha} f'(\Rg) + \squareg \left[ \Rh f''(\Rg) \right]  - \frac { \Rh f'(\Rg) } {2} \right\} \\
&+ h_{\mu \nu} \left[ \squareg f'(\Rg) - \frac {f(\Rg)} {2} \right] + \Gammah^\alpha{}_{\mu \nu} \nablag_{\alpha} f'(\Rg) - \nablag_{\mu} \nablag_{\nu} \left[ \Rh f''(\Rg) \right]  = 8 \pi \Th_{\mu \nu} + \mathcal{O}(h^2) ,
\end{aligned}
\end{equation}
where the $\Gamma^{i}_{jk}$ form the Christoffel symbols. Equations \eqref{eq:generalsystem} are complicated, and although one can write them down explicitly without too much difficulty, they present themselves in a rather unwieldy form since they are coupled and are (generally) fourth-order in $\boldsymbol{h}$. If we specialise to the case of a vacuum Einstein background with $\Rg_{\mu \nu} = 0$, \eqref{eq:generalsystem} simplifies considerably, though is still coupled (i.e. it is hard to disentangle the components of $\boldsymbol{h}$ from each other) and presents a challenging numerical problem.

%



\section{Perturbation equations in coordinate form for practical use}

For a Petrov D, vacuum Einstein background, the Newman-Penrose formalism can be used to reduce the complexity of the general system \eqref{eq:generalsystem}, as  demonstrated in \cite{suv19}. The major advantage of this latter work was that the metric was expanded in such a way that, in an appropriate gauge, the resulting equations decouple, as in GR \cite{teu73}. In particular, the relevant equations governing gravitational perturbations of a Kerr black hole in $f(R)$ gravity are given (in the NP language) by expressions (33)-(36) of \cite{suv19}. It is, however, useful to present the NP system of equations in coordinate form, so that they can be tackled more easily with numerical algorithms. To this end, it is convenient to introduce the Kinnersley null-tetrad $\{\boldsymbol{\ell}, \boldsymbol{n}, \boldsymbol{m}, \bar{\boldsymbol{m}}\}$ in Boyer-Lindquist coordinates $\{t,r,\theta,\phi\}$ \cite{kinn69},
\begin{equation} \label{eq:ellkerr}
\ell^{\mu} = \frac{\left( r^2 + a^2  , 1 ,0 , a  \right)} {\Delta}, \,\,\, n^{\mu} = \frac {\left( r^2 + a^2 , - \Delta, 0 ,a \right)} {2  \Sigma}, \,\,\  m^{\mu} = \frac { \left( i a \sin \theta, 0, 1, i \csc\theta \right) } {\sqrt{2} \Sig },
\end{equation}
with $\Delta = r^2 - 2 M r + a^2$, $\Sig = r + i a \cos \theta$, and $\Sigma = \Sig \Sigb = r^2 + a^2 \cos^2\theta$, where $M$ and $a$ represent the mass and the spin parameter of the black hole, respectively. From expressions \eqref{eq:ellkerr}, the Kerr metric is given through the general formula
\begin{equation} \label{eq:kerrmet}
g_{\mu \nu} = \ell_{\mu} n_{\nu} + n_{\mu} \ell_{\nu} - m_{\mu} \mb_{\nu} - \mb_{\mu} m_{\nu}.
\end{equation}

In terms of the Kinnersley tetrad \eqref{eq:ellkerr}, a general metric perturbation \eqref{eq:mettpt} can now be written
\begin{equation} \label{eq:pertmetexp}
\begin{aligned}
h_{\mu \nu} =& h_{nn} \ell_{\mu} \ell_{\nu} - 2 h_{ n \mb} \ell_{ ( \mu} m_{ \nu )} - 2 h_{n m} \ell_{(\mu} \mb_{\nu )} + 2 h_{\ell n} \ell_{( \mu} n_{\nu)} + h_{\ell \ell} n_{\mu} n_{\nu} - 2 h_{\ell \mb} n_{(\mu} m_{\nu)} \\
&- 2 h_{\ell m} n_{(\mu} \mb_{\nu)}  + h_{mm} \mb_{\mu} \mb_{\nu} + 2 h_{m \mb} m_{(\mu} \mb_{\nu )} + h_{\mb \mb} m_{\mu} m_{\nu} ,
\end{aligned}
\end{equation}
where the round brackets denote the symmetrisation operation: $X_{(ij)k} \equiv \tfrac{1}{2} \left( X_{ijk} + X_{jik} \right)$. To simplify the general expression \eqref{eq:pertmetexp}, we employ the so-called outgoing radiation gauge $\ell^{\mu} h_{\mu \nu} = 0$ (see \cite{price07} for a discussion on the existence of this gauge choice). Additionally, noting that
\begin{equation} \label{eq:conj1}
\overbar{h_{m m}} \equiv \overbar{ \left( m^{\mu} m^{\nu} h_{\mu \nu} \right)} = \mb^{\mu} \mb^{\nu} h_{\mu \nu} \equiv h_{ \mb \mb},
\end{equation}
and
\begin{equation} \label{eq:conj2}
\overbar{h_{n m}} \equiv \overbar{ \left( n^{\mu} m^{\nu} h_{\mu \nu} \right)} = n^{\mu} \mb^{\nu} h_{\mu \nu} \equiv h_{ n \mb},
\end{equation}
implies that solving for the functions $h_{nn}, h_{nm}, h_{m \mb}$, and $h_{m m}$, would be enough to completely reconstruct the perturbed metric \eqref{eq:pertmetexp}.

Using the results obtained in \cite{suv19}, we can now write down the decoupled equations describing an oscillating Kerr black hole in $f(R)$ gravity. For simplicity, we consider vacuum perturbations $\Th_{\mu \nu} = 0$, and do not write down the Teukolsky equations (which are the same in vacuo for GR and $f(R)$ gravity) describing the evolution of the Weyl scalars $\psih$; these can be found (in coordinate form) in \cite{teu73}. The full set of perturbation equations, presented in an order which, when solved sequentially, results in a decoupled system, read

\begin{equation} \label{eq:r0}
\begin{aligned}
&3 a_{2} \Bigg\{ \left[ \frac {\left(r^2 + a^2 \right)^{2}} {\Delta} - a^2 \sin^2\theta \right] \partial_{t}^{2} + \frac {4 M a r} {\Delta} \partial_{t} \partial_{\phi} + \left( \frac {a^2} {\Delta} - \csc^2\theta \right) \partial_{\phi}^{2} \\& - \partial_{r} \left( \Delta \partial_{r} \right) - \csc\theta \partial_{\theta} \left( \sin\theta \partial_{\theta} \right)  \Bigg\} \Rh - \Sigma \Rh = 0,
\end{aligned}
\end{equation}

\begin{equation} \label{eq:h1}
\hspace{-2cm}\begin{aligned}
&\Big\{ \left(r^2 + a^2 \right) \left[ \left(r^2 + a^2 \right) \partial_{t}  + 2 \Delta \partial_{r} + 2 a \partial_{\phi} + \left( \frac {4a^2} {a^2+r^2} - 2 \right) M + \frac {2 r \Delta} {\Sigma} \right] \partial_{t} + \Delta \left[ \Delta \partial_{r}  + 2 a \partial_{\phi} + \frac {2 r \Delta} {\Sigma} \right] \partial_{r} \\
&+ a^2 \partial_{\phi}^2 + \frac {2 a r \left(a^2 - M r \right) + 2 a^3 \left( M - r \right) \cos^2\theta} {\Sigma} \partial_{\phi}  + \frac {4 a^2 \Delta^2 \cos^2\theta} {\Sigma^2} \Big\} h_{m \mb} =  S_{1},
\end{aligned}
\end{equation}
\begin{equation}
\hspace{-2cm}\begin{aligned}
&\Big\{ \left( r^2 + a^2 \right) \left[ \left( r^2 + a^2 \right) \partial_{t}  + 2 \Delta \partial_{r} +2 a \partial_{\phi} + \left(\frac {4a^2} {r^2 + a^2} - 2 \right) M + \frac {2 \Delta} {\Sig} \right] \partial_{t} + \Delta \left[ \Delta \partial_{r} + 2 a \partial_{\phi} + \frac {2 \Delta} {\Sig} \right] \partial_{r} \\
&+ a^2 \partial_{\phi}^2 + \frac {2 a \left[ a^2 - M r + i a (M - r) \cos\theta \right]} {\Sig} \partial_{\phi} \Big\} h_{mm} =  2 \Delta^2 \psih_{0},
\end{aligned}
\end{equation}
\begin{equation}
\hspace{-2cm}\begin{aligned}
&\Big\{ \left( r^2 + a^2 \right) \left[ \left( r^2 + a^2 \right) \partial_{t}  + 2 \Delta \partial_{r} +2 a \partial_{\phi} + \left(\frac {4a^2} {r^2 + a^2} - 2 \right) M + \frac {2 \Delta} {\Sig} \right] \partial_{t} + \Delta \left[ \Delta \partial_{r} + 2 a \partial_{\phi} + \frac {2 \Delta} {\Sig} \right] \partial_{r} \\
&+ a^2 \partial_{\phi}^2 + \frac {2 a \left[ a^2 - M r + i a (M - r) \cos\theta \right]} {\Sig} \partial_{\phi} + \frac {2 \Delta^2 \left(3 a^2 \cos^2\theta - r^2 \right)} {\Sigma^2} \Big\} h_{n \mb} = S_{31} + S_{32} + S_{33} + S_{34},
\end{aligned}
\end{equation}
and finally
\begin{equation}
\hspace{-2cm}\begin{aligned}
&\Big\{ \left[ - \frac {a^2 \sin^2 \theta} {2} \partial_{t} - i a \sin\theta \partial_{\theta} - a \partial_{\phi}  + \frac {a^2 \left( r + 3 i a \cos\theta \right) \sin^2\theta} {\Sigma} \right] \partial_{t} + \frac{1}{2} \partial_{\theta}^{2} + \left[ \frac {a \left(i r - 3 a \cos\theta \right) \sin\theta} {\Sigma} - \frac {\cot\theta} {2} \right] \partial_{\theta} \\
&- i \csc\theta \partial_{\theta} \partial_{\phi} - \frac {1}{2} \csc^2\theta \partial_{\phi}^{2} + \frac {\left[ i \left(3 a^2 + 2 r^2 \right) \cos\theta + a \left( r - r \cos 2 \theta - i a \cos 3 \theta \right) \right] \csc^2\theta} {2 \Sigma} \partial_{\phi} \\
&+ \frac {a^2 \left(r - 3 i a \cos\theta \right) \sin^2\theta} {\Sigma \Sigb} \Big\} h_{nn} = S_{41} + S_{42} + 2 \Sigb^2 \psih_{4},
\end{aligned}
\end{equation}
where equation \eqref{eq:r0} arises from the trace constraint \eqref{eq:treqn} and the $S$ form source terms. In particular, we have
\begin{equation}
\hspace{-2cm}\begin{aligned}
S_{1} &= -a_{2} \Big\{ \left( r^2 + a^2 \right) \left[ \left( r^2 + a^2 \right) \partial_{t} + 2 \Delta \partial_{r} + 2 a \partial_{\phi} + \frac {2 M \left(a^2 - r^2 \right)}{ r^2 + a^2} \right] \partial_{t} \\
&+ \Delta^2 \partial_{r}^2 + 2 a \Delta \partial_{r} \partial_{\phi} + a^2 \partial_{\phi}^{2} + 2 a \left(M - r \right) \partial_{\phi} \Big\} \Rh ,
\end{aligned}
\end{equation}
\begin{equation}
\hspace{-2cm}\begin{aligned}
S_{31} &= - \frac {\Delta} { \sqrt{2} \Sigb} \Bigg\{ \Big\{ - i a \left( r^2 + a^2 \right) \sin\theta \partial_{t} - i a \Delta \sin\theta \partial_{r} + \left(r^2 + a^2 \right) \partial_{\theta} - i \left[a^2 + \left( r^2 + a^2 \right) \csc^2\theta \right] \sin\theta \partial_{\phi} \\
&+ \frac {a \left[ - i r \left( 3 a^2 + 3 r^2 + 2 M r \right) + a \left( 5 a^2 + 5 r^2 - 6 M r \right) \cos \theta \right] \sin \theta} {\Sigma} \Big\} \partial_{t} \\
&+ \left[\Delta \partial_{\theta} + - i \Delta \csc\theta  \partial_{\phi} + \frac {2 a \Delta \left(a \cos\theta - 2 i r \right) \sin\theta} {\Sigma} \right] \partial_{r} + \left( a \partial_{\phi} + \frac {\Delta} {\Sigb} \right) \partial_{\theta} \\
&+ \frac {\left[ \left( 7 a^3 + 6 r^2 a - 12 M r a \right) \cos\theta + 2 i r \left(r^2 + 2 a^2 \cos 2 \theta - a^2 - 2 M r \right) - a^3 \cos 3 \theta \right] \csc\theta} {2 \Sigma} \partial_{\phi}  \\
&- i a \csc\theta \partial_{\phi}^{2} - \frac {4 i a \Delta \sin\theta} {\Sigma} \Bigg\} h_{m \mb} ,
\end{aligned}
\end{equation}
\begin{equation}
S_{32} = - \frac {\Delta \cot \theta} {\sqrt{2} \Sig} \left[ \left(r^2 + a^2 \right) \partial_{t} + \Delta \partial_{r} + a \partial_{\phi} - \frac {2 i a \Delta \cos \theta} {\Sigma} \right] h_{mm}  ,
\end{equation}
\begin{equation}
\hspace{-2cm}\begin{aligned}
S_{33} &= -\frac {i \Delta \sin \theta} { \sqrt{2} \Sig} \Bigg\{ \Big\{ a \left(r^2 + a^2 \right) \partial_{t} + a \Delta \partial_{r} - i \left(r^2 + a^2 \right) \csc\theta \partial_{\theta} + \left[ a^2 + \left( r^2 + a^2 \right) \csc^2\theta \right] \partial_{\phi} 
\\
& - \frac {a r \left( \Delta + 2 i a M \cos\theta \right) + i \left( r^2 + a^2 \right)^2 \cot\theta \csc\theta} {\Sigma} \Big\} \partial_{t} \\
&+ \Delta \left[ - i \csc\theta \partial_{\theta} +  \csc^2 \theta \partial_{\phi} + \frac {i \left(-2a^2 - r^2 + a^2 \cos^2\theta \right) \cot\theta \csc\theta} {\Sigma} \right] \partial_{r} \\
& + \left( \frac {i \Delta \csc\theta} {\Sig} - i a \csc\theta \partial_{\phi} \right) \partial_{\theta} + a \csc^2 \theta \partial_{\phi}^{2} + \frac {\left[ i a^3 \cos 3 \theta - i a \left(a^2 + 8 M r \right) \cos\theta - 4 r \Delta \right] \csc\theta} {4 \Sigma} \partial_{\phi} \\
&+ \frac {a \Delta \left[ 2 \left(a^2 + 2 r^2 \right) \cos 2 \theta - a^2 \left( \cos 4 \theta - 3 \right) \right] \csc^2\theta} {2 \Sigma^2} \Bigg\} h_{\mb \mb} ,
\end{aligned}
\end{equation}
\begin{equation}
\hspace{-2cm}\begin{aligned}
S_{34} &= - a_{2} \frac {\sqrt{2} i \Delta} {\Sigb} \Bigg\{ \Big\{ - a \left( r^2 + a^2 \right) \sin\theta \partial_{t} - a \Delta \sin\theta \partial_{r} - i \left(r^2 + a^2 \right) \partial_{\theta} - \left[ a^2 + \left(r^2 + a^2 \right) \csc^2\theta \right] \sin\theta \partial_{\phi} \\
& - \frac {2 a M r \sin\theta}{ \Sigb} \Big\} \partial_{t} - \Delta \left( i \partial_{\theta} + \csc\theta \partial_{\phi} + \frac {a \sin\theta} {\Sigb} \right) \partial_{r}  + \frac {i \Delta} {\Sigb} \partial_{\theta} - i a \partial_{\theta} \partial_{\phi} - \csc\theta \left( a \partial_{\phi} - \frac {\Sigma - 2 M r} {\Sigb} \right) \partial_{\phi} \Bigg\} \Rh,
\end{aligned}
\end{equation}
\begin{equation}
\hspace{-2cm}\begin{aligned}
S_{41} &= - \frac {\sin\theta} {\sqrt{2} \Sig} \Bigg\{ \Big\{ i a \left(r^2 + a^2 \right) \partial_{t} - i a \Delta \partial_{r} - \left(r^2 + a^2 \right) \csc\theta \partial_{\theta} + i \left[ a^2 + \left(r^2 + a^2 \right) \csc^2\theta \right] \partial_{\phi} \\
& + \frac { - 2 a i r \left(a^2 + \Delta \right) + a^2 \cos\theta \left[ 4 a^2 - 2 M r + 4r^2 - 2 i a \left( M -r \right) \cos\theta +  \left(r^2 + a^2 \right) \cot^2\theta \right] } {\Sigma} \\
&+ \frac{r^2 \left(r^2 + a^2 \right) \cot\theta \csc\theta } {\Sigma} \Big\} \partial_{t} \\
& + \Delta \left[ \csc\theta \partial_{\theta} - i \csc^2\theta \partial_{\phi} + \frac {i a r + \left( a^2 \cos2\theta - r^2 - 2 a^2 \right) \cot\theta \csc\theta} {\Sigma} \right] \partial_{r} - a \csc\theta \partial_{\theta} \partial_{\phi} \\
&+ \frac {i a \left[ \Delta + 2 i a \left( r - M \right) \cos\theta \right] \cot \theta + r \left( 3 a^2 - 4 M r + r^2 \right) \csc\theta} {\Sigma} \partial_{\theta}  + i a \csc^2\theta \partial_{\phi}^{2} \\
&- \frac {i r \left(4 a^2 - 4 M r + r^2 \right) + a \cos\theta \left[ -3a^2 +2 \left( M - r \right) r + i a \left( 2 M -3 r \right) \cos\theta + a^2 \cos 2 \theta \right]} {\Sigma} \csc^2\theta \partial_{\phi} \\
&+ \frac{1} {8 \Sigma^{2}}  \Big\{ -2 \left[ 3 a^2 r^3 + 4 r^4 \left( -4 M + r \right) + a^4 \left( 8 M + r \right) \right] \cot\theta - i a \csc\theta \Big\{ \left(a^2 - r^2 \right) \left( 7 a^2 - 16 r M + 4 r^2 \right) \\
&+4 \left[ a^4 - 2 a^2 \left( M - 3 r \right) r + r^3 \left( -8 M + 3 r \right) \right] \cos2\theta - 2 i a \left[ a^2 \left( M + 2 r \right) + r^2 \left( 5 r - 8 M \right) \right] \cos 3 \theta \\
&+ a^2 \left( -3 a^2 + 8 M r - 5 r^2 \right) \cos 4 \theta + 2 i a^3 \left( M - r \right) \cos 5 \theta  \Big\} \Big\} \Bigg\} h_{n \mb} ,
\end{aligned}
\end{equation}
and
\begin{equation}
\hspace{-2cm}\begin{aligned}
S_{42} &= -\frac {r^2 + a^2} {2 \Sig^2} \Bigg\{ \big\{ \frac {r^2 + a^2} {2} \partial_{t} - \Delta \partial_{r} + a \partial_{\phi} + \left[ M - \frac {2 a^2 M } {r^2 + a^2} - \frac {2 \Delta} {\Sigb} + \frac {\Delta} {\Sig} \right] \big\} \partial_{t} \\
& + \frac {\Delta^2} {r^2 + a^2} \left[ \frac {1} {2} \partial_{r}  - \frac {a} {\Delta} \partial_{\phi} + \frac {r + 3 i a \cos \theta} {\Sigma} \right] \partial_{r} \\
&+ \frac {a} {r^2 + a^2} \Big\{ \frac {a} {2} \partial_{\phi} + \frac { M r^2 - a^2 r + a \cos \theta \left[ -3 i \Delta + a \left( r - M \right) \cos \theta \right]} {\Sigma} \Big\} \partial_{\phi} - \frac {4a^2 \Delta^2 \cos^2\theta} {\left(r^2 + a^2 \right) \Sigma^2} \Bigg\} h_{\mb \mb}
\end{aligned}
\end{equation}

which, when solved subject to appropriate boundary conditions, defines the gravitational perturbations of Kerr black holes in $f(R)$ gravity. An explicit solution will be presented elsewhere.

\section*{Acknowledgements}
This work was supported by the Alexander von Humboldt Foundation.


\end{document}